\documentclass[conference]{IEEEtran}
\usepackage{amsmath,amssymb,amsfonts}
\usepackage{cite}
\usepackage{graphicx}
\usepackage{psfrag}
\usepackage{subfigure}
\usepackage{amsmath}
\usepackage{array}
\usepackage{algorithmicx}
\usepackage{algpseudocode}
\usepackage{setspace}
\usepackage{blindtext}
\usepackage{url}
\usepackage{epstopdf}
\usepackage{verbatim}
\usepackage{color}
\usepackage{amsmath}
\usepackage{bm}
\usepackage{multirow}
\usepackage[table,xcdraw]{xcolor}
\usepackage{booktabs}
\usepackage{makecell}
\usepackage{ntheorem}
\usepackage{textcomp}
\usepackage{amsmath}

\newtheorem{Prob}{Problem}

\newcommand{\RNum}[1]{\uppercase\expandafter{\romannumeral #1\relax}}

\theoremseparator{:}

\title{High Precision Indoor Localization with Dummy Antennas - An Experimental Study}
\author{Kaixuan Huang$^{\dagger}$, Chenlu Xiang$^{\dagger}$, Shunqing Zhang$^{\dagger}$, Shugong Xu$^{\dagger}$, Xianfeng Ma$^{\ddag}$, Qinglong Xian$^{\ddag}$, 
and Hua Yang$^{\ddag}$\\
$^{\dagger}$Shanghai Institute for Advanced Communication and Data Science, Shanghai University, Shanghai, China\\
$^{\ddag}$Guangzhou Guangzhong Enterprise Group Corporation, Guangdong, China\\
Email: \{kaixuanhuang, xcl, shunqing, shugong\}@shu.edu.cn,\{mxf,xian.ql,08250\}@gzhm.com}

\begin{document}
\maketitle
\begin{abstract}
With the rising demand for indoor localization, high precision technique-based fingerprints became increasingly important nowadays. The newest advanced localization system makes effort to improve localization accuracy in the time or frequency domain, for example, the UWB localization technique can achieve centimeter-level accuracy but have a high cost. Therefore, we present a spatial domain extension-based scheme with low cost and verify the effectiveness of antennas extension in localization accuracy. In this paper, we achieve sub-meter level localization accuracy using a single AP by extending three radio links of the modified laptops to more antennas. Moreover, the experimental results show that the localization performance is superior as the number of antennas increases with the help of spatial domain extension and angular domain assisted. 
\end{abstract}


\maketitle
\section{Introduction} \label{sect:intro}
Location-based services (LBS) become a key element of modern mobile internet applications, which utilize real-time geometric information from smartphones to provide navigation, entertainment, or security \cite{overview}. Although satellite positioning systems, such as global positioning system (GPS) \cite{GPS} or Beidou \cite{Jiang2018Beidou}, can achieve centimeter-level localization accuracy in the outdoor environment, they can hardly achieve the same level in the indoor scenario, due to satellite signal occlusions. To address this issue, more diversified wireless signals, including wireless fidelity (WiFi)\cite{wifi1}, Bluetooth low energy (BLE)\cite{ble}, and increasingly popular 3GPP long term evolution (LTE) / new radio (NR) technology \cite{LTE}, have been exploited to provide indoor localization services, and accurate navigation and seamless tracking capability in both outdoor and indoor scenarios have been specified in 3GPP Release 16 standard \cite{3GPP}.

In order to provide a better indoor localization accuracy, the most direct approach is to collect more observation samples, either in the frequency or time domain \cite{vasisht2016decimeter}. For example, the ultra wide band (UWB) localization technique adopted in the recent iPhone product, is able to achieve centimeter-level localization accuracy with 500 MHz bands. Fingerprint-based localization solutions can also achieve a similar accuracy when the observation window is sufficiently long \cite{8409950}. However, improving the localization accuracy via spatial domain extension is never straightforward due to the following reasons. First, the existing spatial domain schemes \cite{RBF2010,TOA} rely on calculating the geometric location through the measured time of flight (TOF) and angle of arrival (AoA), and the corresponding resolution does not scale linearly with respect to the number of available radio frequency (RF) chains \cite{2014CSIMIMO}. Second, deploying more RF chains with expensive components, including low-noise amplifiers, analog-to-digital, or digital-to-analog converters, are cost-prohibitive in general \cite{SWAN}. Third, the processing complexity increases exponentially with respect to the number of RF chains as well \cite{Flexcore}. As illustrated in \cite{2013BigStation}, the baseband processing of a 12-antenna software defined radio (SDR) system has to be moved to `BigStation' with 15 personal computers for real-time signal processing.

Last but not least, to address the aforementioned issues of the spatial domain extension for high accuracy indoor localization, we present in this paper a low-cost WiFi based {\bf {\em prototype}} localization system with dummy antennas to improve AoA resolutions as well as localization accuracy. By regularly rotating 12 antennas with only 3 RF chains, the proposed system is able to achieve {\bf less than 1 meter} localization accuracy in a $40 m^{2}$ corridor environment, and the main contributions of our work are listed below.
\begin{itemize}
\item{\em Low-Cost Spatial Domain Extension} Conventional spatial domain extension schemes require special hardware support. As reported in \cite{2014Phaser}, Phaser builds a 9-antenna WiFi based AoA estimation system by combining multiple WiFi network interface cards (NICs) and the associated RF chains together, and the hardware cost is thus significant. In this paper, we control the number of RF chains and extend the number of antennas by introducing a four-port switcher to reduce the hardware cost, which can enjoy the benefit of AoA resolution improvement and maintains the hardware cost simultaneously.
\item{\em AoA Assisted Localization} Different from the conventional approach to directly minimize mean square error (MSE) in \cite{2019xiang}, we propose a two-stage localization strategy, which contains an AoA estimation and a location prediction. Through this approach, the localization accuracy can be improved to less than 1 meter with more available dummy antennas.
\item{\em Low Cost Hardware Implementation} We use three low-cost off-the-shelf single pole four throws (SP4T) RF switchers to connect 12 dummy antennas on top of only one 3-port WiFi NIC. By applying a time division based activation scheme, we are able to collect the channel behaviors of all the available antennas, which eventually pave a way for low-cost high precision indoor localization with dummy antennas. 
\end{itemize}
 
The remainder of this paper is organized as follows. In Section~\ref{sect:pre}, we introduce the system model and some preliminary information, and the problem formulation is discussed in Section~\ref{sect:prob}. The problem analysis, as well as the corresponding solution, are provided in Section~\ref{sect:solution}. In Section~\ref{sect:experiment}, we present our experimental results and the concluding remarks are given in Section~\ref{sect:conc}.

\section{System Model} \label{sect:pre}
In this section, we briefly introduce some preliminary knowledge about channel state information (CSI), followed by the angular based localization procedures.

\subsection{CSI Modeling}
Consider an orthogonal frequency division multiplexing (OFDM) based modern WiFi communication system with $N_{SC}$ subcarriers. For a given location $\mathcal{L}$, the received signal vector, $\mathbf{y}(\mathcal{L}) = \left[y_1(\mathcal{L}), \cdots, y_{N_R}(\mathcal{L}) \right]^T \in \mathbb{C}^{N_R \times 1} $, can be modeled through,
\begin{eqnarray}
\mathbf{y}(\mathcal{L}) = \mathbf{H}(\mathcal{L}) \cdot \mathbf{x} + \mathbf{n}(\mathcal{L}),
\end{eqnarray}
where $\mathbf{H}(\mathcal{L}) \in \mathbb{C}^{N_R \times N_{SC}}$ denotes the channel frequency responses, $\mathbf{x} \in \mathbb{C}^{N_{SC} \times 1}$ denotes the transmitted symbols, and $\mathbf{n}(\mathcal{L}) \in \mathbb{C}^{N_R \times 1}$ denotes the additive white Gaussian noise (AWGN), respectively. If we further choose $h_i(\mathcal{L}, n)$ to be the collected channel state information of the $i^{th}$ subcarrier at the $n^{th}$ received antenna for a given location $\mathcal{L}$, $\mathbf{H}(\mathcal{L})$ can be rewritten as,
\begin{eqnarray}
\mathbf{H}(\mathcal{L}) = \left[
\begin{array}{ccc}
h_1(\mathcal{L},1) & \cdots & h_{N_{SC}}(\mathcal{L},1) \\
\vdots & \ddots & \vdots \\
h_1(\mathcal{L},N_R) & \cdots & h_{N_{SC}}(\mathcal{L},N_R) 
\end{array}
\right].
\end{eqnarray}

Without loss of generality, we denote $a_i(\mathcal{L}, n)$ and $p_i(\mathcal{L}, n)$ to be the amplitude and phase information of the corresponding channel responses, $h_i(\mathcal{L}, n)$ can be modeled as, 
\begin{eqnarray}
h_i(\mathcal{L}, n) = a_i(\mathcal{L}, n) \cdot e^{j \cdot p_i(\mathcal{L}, n)}.
\end{eqnarray}
\begin{eqnarray}
\mathbf{P}(\mathcal{L}) = \left[
\begin{array}{ccc}
p_1(\mathcal{L},1) & \cdots & p_{N_{SC}}(\mathcal{L},1) \\
\vdots & \ddots & \vdots \\
p_1(\mathcal{L},N_R) & \cdots & p_{N_{SC}}(\mathcal{L},N_R) 
\end{array}
\right].
\end{eqnarray}
and thus we obtain phase matrix $\mathbf{P}(\mathcal{L})$ extracted by $\mathbf{H}(\mathcal{L})$.                 

\subsection{AoA Assisted Localization}
\begin{figure}
\centering
\includegraphics[width = 3.4 in]{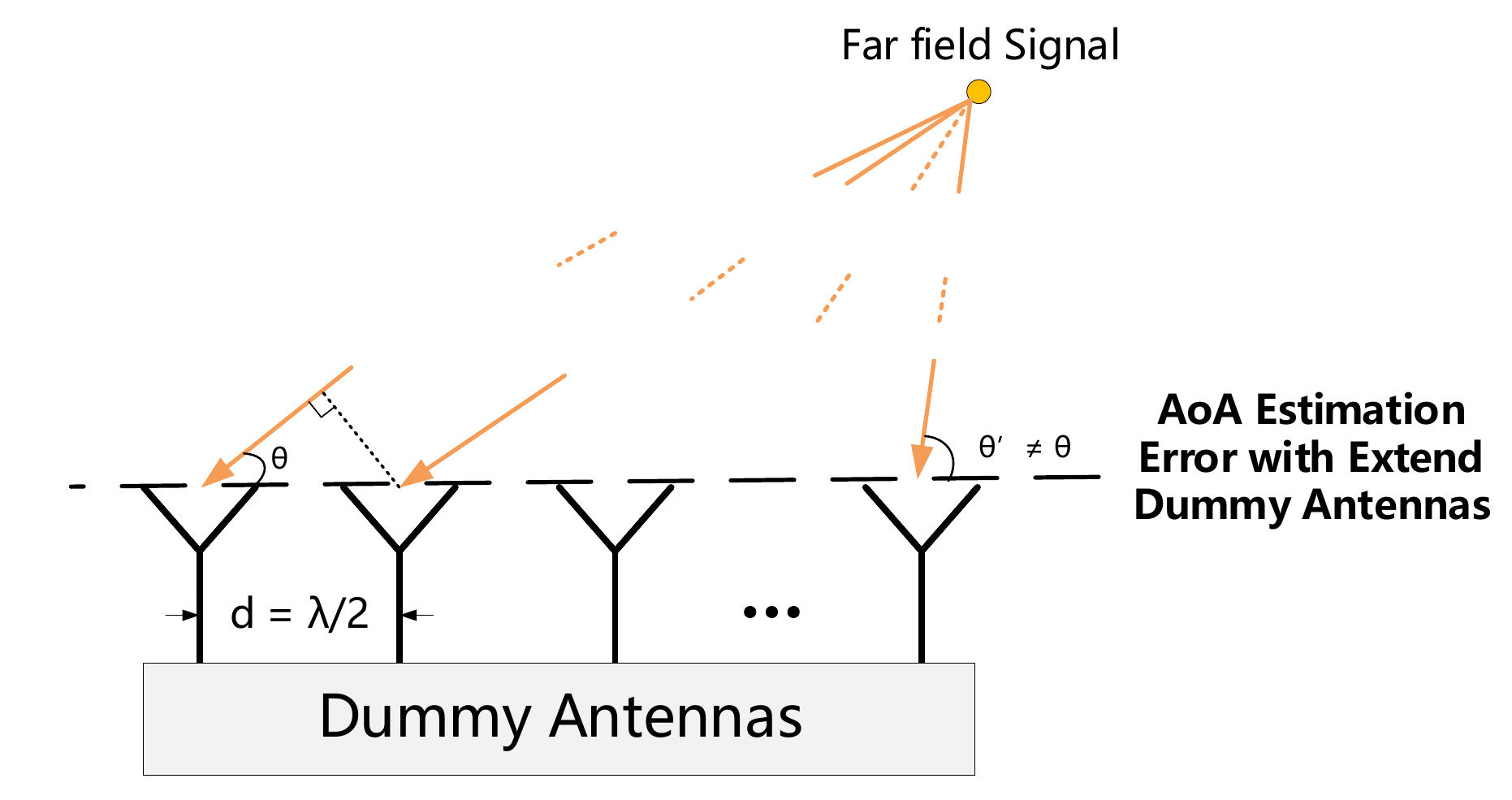}
\caption{Illustration diagram for AoA estimation with multiple receiving antennas. With extended dummy antennas, the AoA estimation becomes inaccurate.}
\label{fig:findAOA}
\end{figure}

As illustrated in Fig.~\ref{fig:findAOA}, the geometric distances between two adjacent antennas are selected to be half of the wavelength, and therefore, the AoA of uniform linear array (ULA) configuration can be estimated via equation \eqref{findaoa},
\begin{eqnarray}
\bar{\theta} = \frac{1}{N_{SC}} \sum_{i=1}^{N_{SC}} \arccos\left(\frac{p_i(\mathcal{L}, n+1) - p_i(\mathcal{L}, n)}{\pi}\right).
 \label{findaoa}
\end{eqnarray}

However,  the measured CSI usually contains information of the non-line-of-sight (NLOS) path, in practical, which makes the AoA estimation inaccurate, equation~\eqref{findaoa} does not work directly. In order to mitigate the effect of NLOS, we need to extend the number of the antenna array to estimate AoA more accurately. When the number of antennas is greater than the number of NLOS paths, the MUSIC algorithm could be employed to estimate the AOA of the multi-path signal.

Considering a ULA with $N_R$ antennas, the correlation matrix of received signals vector $\mathbf{R}_{\mathbf{yy}}$ can be expressed by,
\begin{eqnarray}
\mathbf{R}_{\mathbf{yy}} = \mathbb{E}[\mathbf{yy}^H]
\end{eqnarray}
where $\mathbb{E}(\cdot)$ and $(\cdot)^H $ denote conjugate transpose and expectation, respectively. 
The singular value decomposition of $\mathbf{R}_{\mathbf{yy}}$ is performed, and obtain $N_R$ eigenvalues corresponding to $N_R$ vectors $\mathbf{U} = [u_1, u_2, \cdots, u_{N_R}]$, which is divided to two parts given as,
\begin{eqnarray}
\mathbf{U} = [\mathbf{U_S} ,\mathbf{U_N}] = [\underbrace{u_1, \cdots, u_p}, \underbrace{u_{p+1}, \cdots, u_{N_R}}]
\end{eqnarray}            
where $\mathbf{U_S}$ and $\mathbf{U_N}$ are the corresponding vectors of signal and noise parts. The signal steering vector $a(\theta_{x}) = [1, e^{-j\cdot w(2, \theta_{x})}, \cdots, e^{-j\cdot w(N_R, \theta_{x})}]^T$ and noise are uncorrelated, we have,
\begin{eqnarray}
a^H(\theta_{x})\mathbf{U_N} = 0 
\end{eqnarray}

Based on the orthogonality based signal and noise, the AoA spectrum of MUSIC can be expressed as,
\begin{eqnarray}
\mathbf{B}(\theta) = \frac{1}{a^H(\theta)\mathbf{U_N}\mathbf{U_N^H}a(\theta)}
\end{eqnarray}

We could obtain the closed expression of steering vector $a(\theta)$ by minimizing its spectrum function, and then AoA of incident signal $\theta_{x}$ can be obtained by spectrum peaking search. We define $\mathcal{M}(\cdot)$ to represent the peak search process of MUSIC algorithm as follows,
\begin{eqnarray}
\theta_{x} = \mathcal{M}\left(\mathbf{B}(\theta)\right)
\end{eqnarray}

Once we obtain the AoA, $\theta_{x}$, the location information could be further derived by combining the distance information.

\section{Problem Formulation}
\label{sect:prob}
In this section, we formulate the localization problem using a generalized optimization framework. Denote $\hat{\mathcal{L}}^{m}$ and $\mathcal{L}^{m}$ to be the $m^{th}$ predicted and true location, respectively, and we choose MSE of them to be the performance measure, where\footnote{$\|\cdot\|_2$ denotes the $l_2$ norm of the inner vector.} $\overline{\Delta\mathcal{L}} = \frac{1}{M}\sum_{m=1}^{M} \|\hat{\mathcal{L}}^{m} - \mathcal{L}^{m}\|_2$. Mathematically, the location error minimization problem can be described as follows.
\begin{Prob}[\em MSE Minimization] The original location error minimization problem is given as follows. \label{prob:mse}
\begin{eqnarray}
\underset{g(\cdot)}{\textrm{minimize}} && \frac{1}{M}\sum_{m=1}^{M} \|\hat{\mathcal{L}}^{m} - \mathcal{L}^{m}\|_2,
\label{eqn:mini1}\\
\textrm{subject to} && \hat{\mathcal{L}}^m = g\left({\widehat{\mathbf{H}}} (\mathcal{L}^m), \bar{\theta}^{m}_{x} \right), \nonumber \\
&&\bar{\theta}^{m}_{x} = \mathcal{M}\left(\mathbf{B}(\theta)\right), \nonumber \\
&& \hat{\mathcal{L}}^m , \mathcal{L}^m \in \mathcal{A}, 
\end{eqnarray}
where $\mathcal{A}$ denotes all the feasible localization areas.
\end{Prob}

In the indoor localization, the accuracy of AoA estimation based on MUSIC algorithm is in general limited, due to the complicated NLOS fading environments. Meanwhile, the associated computational complexity could be significant, specially when the received SNR is small \cite{D-Watch}. To overcome this obstacle, we minimize the mean absolute error (MAE) of AoA rather than directly calculate the AoA value using MUSIC. Denote $\hat{\theta}^m$ and $\theta^m$ to be predicted and true AoA values, we can define the AoA MAE minimization problem as follows.

\begin{Prob}[\em AoA MAE Minimization] The AoA MAE minimization problem can be formulated as follows. \label{prob:aoa}
\begin{eqnarray}
\underset{f(\cdot)}{\textrm{minimize}} && \frac{1}{M}\sum_{m=1}^{M} \|\hat{\theta}^m - \theta^{m}\|,
\label{eqn:mini2}\\
\textrm{subject to} && \hat{\theta}^m = f\left({\widehat{\mathbf{P}}} (\mathcal{L}^m)\right),
\end{eqnarray}
where $\|\cdot\|$ denotes the $l_1$ norm and $\theta^{m}$ is the ground truth value that can be measured from $\mathcal{L}^m$ and the location of AP, $\mathcal{L}^{AP}$.
\end{Prob}

By combining the above two optimization problems together, we propose the following angular domain assisted localization scheme, where we first estimate the AoA value from Problem~\ref{prob:aoa} and then minimize the location error of Problem~\ref{prob:mse} with the estimated AoA value, $\hat{\theta}^{m}$. Mathematically, we can directly model the regression function $g^{\star}(\cdot)$ described as follows. 
\begin{Prob}[\em Angular Domain Assisted Localization] \label{prob:adal} The angular domain assisted localization problem can be formulated as,
\begin{eqnarray}
g^{\star}(\cdot) = \underset{g(\cdot)}{\textrm{minimize}} && \frac{1}{M}\sum_{m=1}^{M} \|\hat{\mathcal{L}}^{m} - \mathcal{L}^{m}\|_2.
\label{eqn:mini3}\\
\textrm{subject to} && \hat{\mathcal{L}}^m = g\left({\widehat{\mathbf{H}}} (\mathcal{L}^m), \hat{\theta}^m \right), \nonumber \\
&&\hat{\theta}^m = f^{\star}\left({\widehat{\mathbf{P}}} (\mathcal{L}^m)\right), \nonumber \\
&& \hat{\mathcal{L}}^m , \mathcal{L}^m \in \mathcal{A},
\end{eqnarray}
where $f^{\star}(\cdot)$ is the optimal solution of Problem~\ref{prob:aoa}.
\end{Prob}

Through the above formulation, we can approximate $f^{\star}(\cdot)$ using the classical logistic regression model in the offline training rather than running MUSIC algorithms in the online stage, which greatly reduces the computational complexity in the localization. Meanwhile, it is also worth mentioning that the localization accuracy and the computational complexities can trade-off with each other in the theoretical sense, when the numbers of training reference points or receive antennas increase, e.g., the localization accuracy can be improved if we increase the experimental cost and algorithm complexity in the offline stage.

\begin{figure}
\centering
\includegraphics[width = 3.4 in]{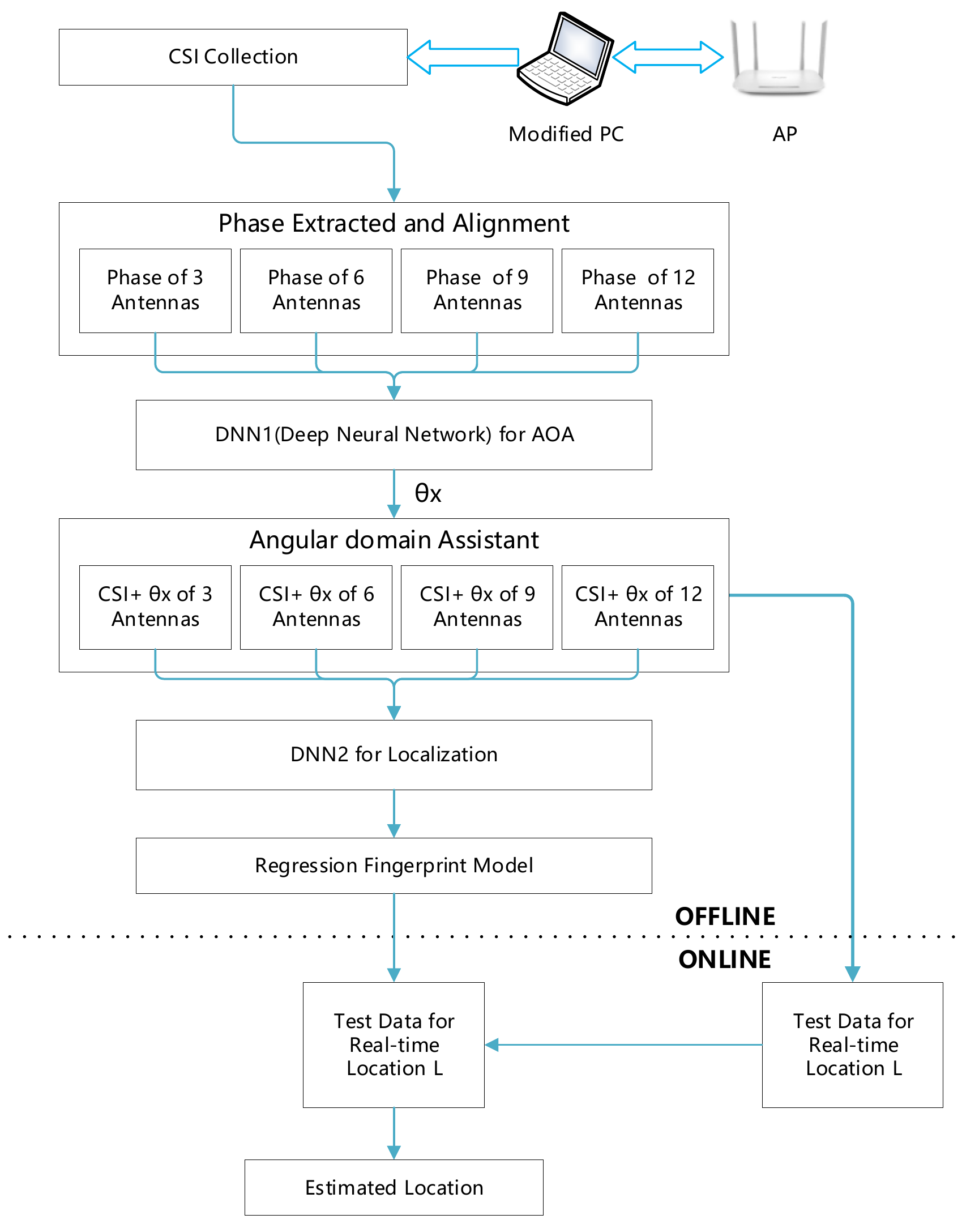}
\caption{The architecture of the system. In the offline stage, we collected CSI of different antennas at Reference Point with modified PC by receiving packets from WiFi AP and the amplitude and phase extracted from collected CSI is the input data of DNN. In offline stage, we collect the CSI at Test Point as test data.}
\label{fig:architecture}
\end{figure}

\section{Deep Learning based Localization Solution}
\label{sect:solution}
In this section, we propose to apply the deep neural network (DNN) architectures to solve the AoA MAE minimization and the angular domain assisted localization as defined in Problem~\ref{prob:aoa} and Problem~\ref{prob:adal}.

\subsection{Phase Alignment}
One of the key factors for applying deep learning based solution is to generate high quality measured CSI. However, the measured phase information usually contains random jitters and noises, due to the imperfect hardware components that used in the practical systems. To solve this issue and obtain high accurate localization results, we model the main phase errors from carrier frequency offset (CFO) and sampling frequency offset (SFO), which is given below.
\begin{eqnarray}
\widehat{p_{i}} = p_{i} + 2\pi\frac{i}{N_{SC}}\Delta{t} +\beta + Z,
\end{eqnarray}
where $\widehat{\angle{p_i}}$ and $\angle{p_i}$ denote the measured and genuine phase of subcarrier $i$, respectively. $\Delta{t}$ is the SFO time lag, $\beta$ is the CFO, and $Z$ is the random measured noise with zero mean and unit variance.

In order to eliminate these random phase offsets, we perform the linear regression \cite{PhaseFi} on the raw phase information across the entire frequency band, where the slope $k$ and the y-intercept of the regression line $b$ can be obtained from, 
\begin{eqnarray}
\label{sect:cali_ph}
p_{i} & = & \widehat{p_{i}} - ki - b, \\
k & = & \frac{\widehat{p_{N_{sc}}}-\widehat{p_1}}{N_{sc}-1}, \\
b & = & \frac{1}{N_{sc}}\sum_{i=1}^{N_{sc}}\widehat{p_i}.
\end{eqnarray}
With the above linear regression, the calibrated phase information can be obtained by subtracting the linear phase offset from the measured phase according to formula \eqref{sect:cali_ph}.

\subsection{Neural Network Configuration}
\label{subsect:NNarchitecture}
\begin{figure}
\centering
\includegraphics[width = 3.4 in]{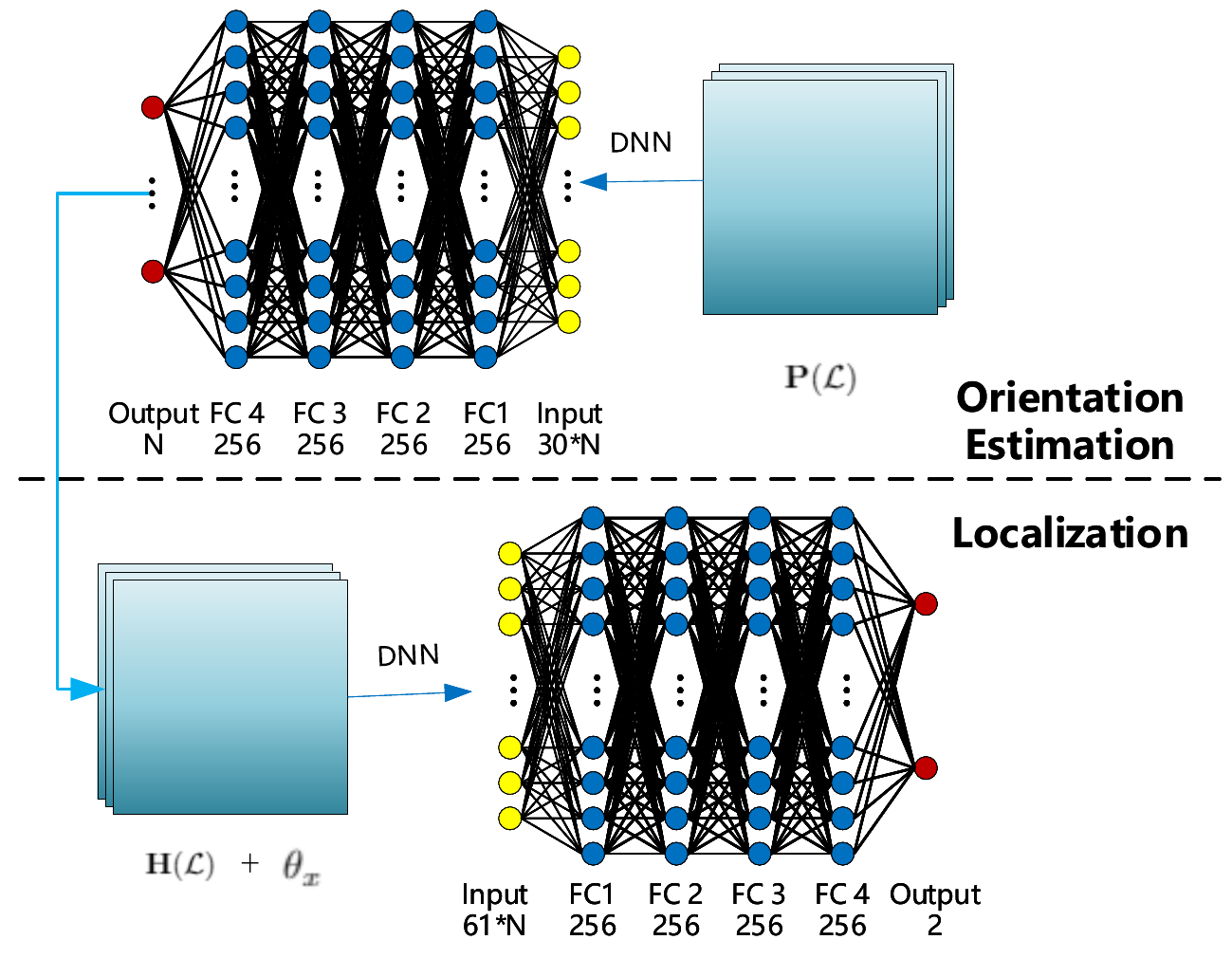}
\caption{The architecture of the DNN, which is consisted of four full connect layers as hidden layer.}
\label{fig:DNN}
\end{figure}

The full connection structure of DNN networks can extract the features from CSI across the spatial domains. To improve the ability to extract features, we implement AoA estimation and localization using two DNNs with four hidden layers, respectively. For two DNNs, we adopt rectified linear unit (ReLU) as the non-linear activation functions in each hidden layer to avoid the vanishing gradient problem and adopt the linear function as the activation function of the last output layer. The architecture of DNN is shown in Fig.~\ref{fig:DNN}.

In DNN1 used for AoA estimation, the input size of the input layer is $30 \times N$, which contains processed phase of 30 subcarriers at $N$ received antennas, the output of DNN1 is AoA. And we select the loss function, $\mathbb{L}_{AoA}$, to be the original definition of $f^*(\cdot)$ expressed as,
\begin{eqnarray}
    \mathbb{L}_{AoA} = \frac{1}{N_{RP}}\sum_{m=1}^{N_{RP}} \|\hat{\theta}^m - \theta^{m}\|, 
\end{eqnarray}

In DNN2 for used localization, every antenna has an additional angle provided by DNN1 and amplitude information of 30 subcarriers, and thus the size of input layer is 61. The output of DNN2 is the coordinate of every RP. Therefore, we select the loss function  $\mathbb{L}_{Loc}$ expressed as,
\begin{eqnarray}
    \mathbb{L}_{Loc} = \frac{1}{N_{RP}}\sum_{m=1}^{N_{RP}} \|\hat{\mathcal{L}}^{m} - \mathcal{L}^{m}\|_2.
\end{eqnarray}

However, in order to avoid the biasing effects of some unusual samples, we adopt the max-pooling technique to get rid of unimportant features and apply the dropout technique to further reduce the unimportant connections in both DNNs. The logistic regression based approach can gradually converge to the non-convex function $\mathbb{L}_{AoA}$ and $\mathbb{L}_{Loc}$ with satisfied performance via respective DNN networks. The detail of the configuration is shown in Table~\ref{tab:parameter}.
\begin{table} [t]
\centering
\caption{An Overview Of Network Configuration and Parameters.}
\label{tab:parameter}
\footnotesize
\renewcommand\arraystretch{1.5}
	\begin{tabular}{c c c}  
	\hline  
	Layers&DNN1&DNN2 \\
	\toprule
	Input Layers&$30\times N$&$61\times N$ \\
	\hline  
    Hidden Layer 1&FC 256 + ReLU&FC 256 + ReLU \\
	\hline  
    Hidden Layer 2&FC 256 + ReLU&FC 256 + ReLU \\
	\hline  
    Hidden Layer 3&FC 256 + ReLU&FC 256 + ReLU \\
	\hline  
    Hidden Layer 4&FC 256 + ReLU &FC 256 + ReLU\\
    & + Dropout 0.3& + Dropout 0.3\\
	\hline  
    Output Layer&FC N + Linear&FC 2 + Linear \\
	\bottomrule		
	\end{tabular}
\end{table}

\section{Experimental Results} \label{sect:experiment}
In this section, we provide some experimental results to validate the proposed high precision indoor localization scheme with 12 dummy antennas. With more receiving antennas installed at the localization terminals, we offer a low cost implementation strategy on top of an off-the-shelf laptop, and compare AoA estimation and localization results with baselines in what follows.

\subsection{Low Cost Implementation}
\begin{figure}
\centering
\includegraphics[width = 3.4 in]{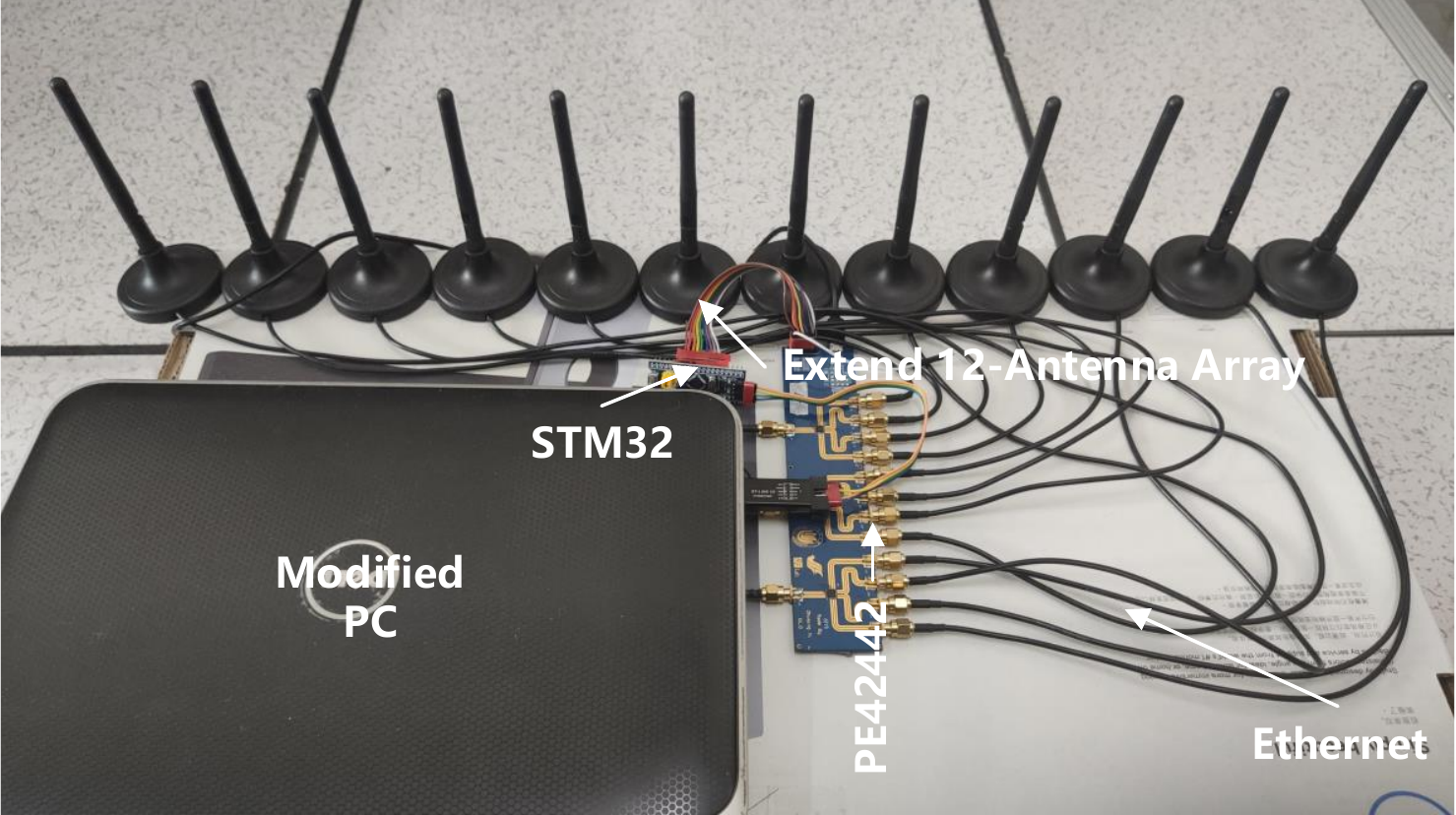}
\caption{The hardware design of extended antenna array. PE42442 is an integrated board of three SP4Ts, which is used to extend 3 radio chains to 12-antenna-array.}
\label{fig:hardware}
\end{figure}

The laptop used in this experiment consists of three receiving antennas. In order to facilitate more receiving antennas, we connect the original three radio chains with 12 antennas via three SP4T RF switches as shown in Fig.~\ref{fig:hardware}. Through this approach, each radio chain can serve four antennas through SP4T RF switches via a time division multiplexing mode, and an external micro-controller unit is applied to control the active durations for each antenna. 

\begin{figure}
\centering
\includegraphics[width = 3.4 in]{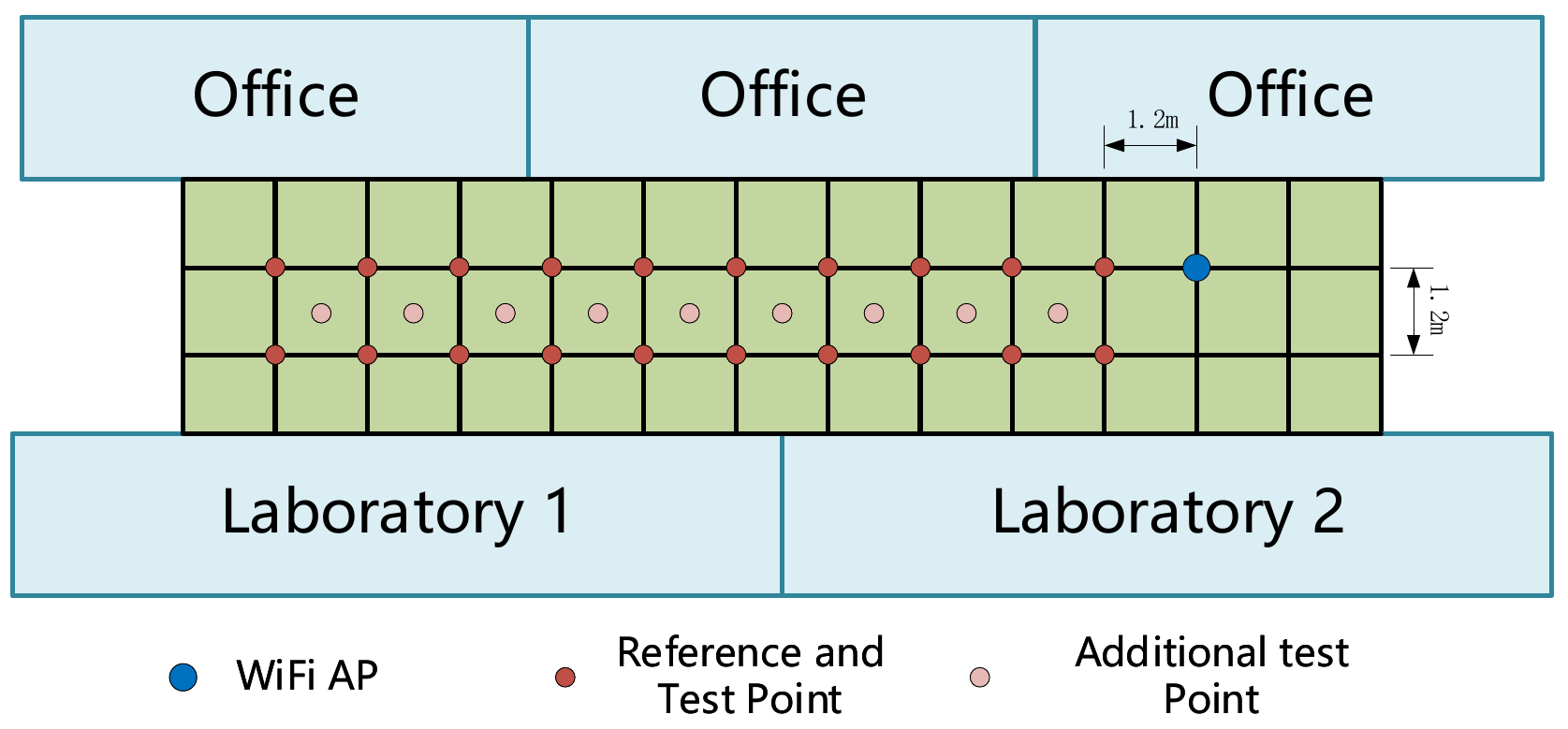}
\caption{A sketch map of the experiment environment. Blue, red and yellow points represent WiFi AP, reference point, and additional test point, respectively.}
\label{fig:scenario}
\end{figure}

To implement localization service, we select a $1.2 \times 12 m^2$ corridor environment with 20 reference points (RPs) and a single WiFi AP. The entire layout of the experimental environment is shown in Fig.~\ref{fig:scenario}, where we uniformly select 9 additional testing points to obtain more reliable results. The 12-antenna modified laptop communicates with WiFi AP to obtain the CSI information through Linux-802.11n-CSI-Tool \cite{2010Predictable} and WiFi network interface cards (NICs). Since we can not concurrently measure all 12 antennas, we divide them into four groups, and collect the corresponding CSIs by fast switching and matching.

During each receiving period, we obtain one CSI packet per antenna, which contains a total of 30 subcarriers. In the testing stage, we collect $20000$ CSI packets per RP for each antenna to construct the training data set, while in the validation stage, $1000$ CSI packets per testing point (TP) for each antenna is applied.  

\subsection{AoA Estimation Results}
\begin{figure}
\centering
\includegraphics[width = 3.4 in]{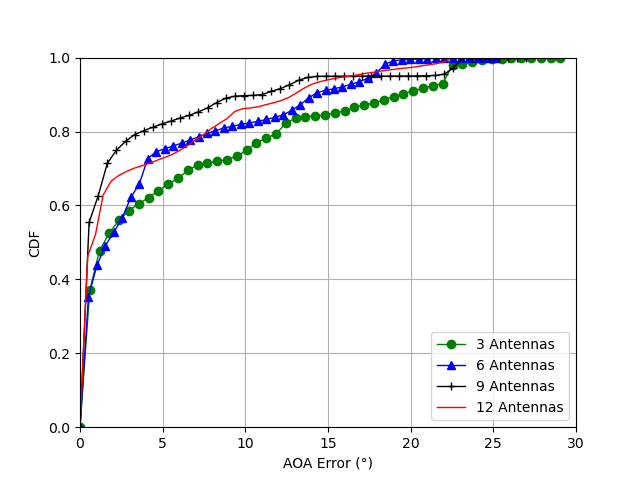}
\caption{The CDF of AoA errors for different number of antennas.}
\label{fig:AOA}
\end{figure}

In the first experiment, we compare the accuracy of AoA estimation for different antenna configurations. The cumulative distribution functions (CDFs) of absolute AoA errors for 3, 6, 9, and 12 receiving antennas are plotted in Fig.~\ref{fig:AOA} and the AoA MAE values are summarized in Table~\ref{tab:MSE}.

From the above results, we find that the AoA MAE value decreases when the number of receiving antennas increases from 3 to 9 antennas. The physical interpretation is that the performance of AoA estimation can be improved, when the number of receiving antennas increases. However, when the number of receiving antennas is greater than 12, the AoA estimation error may become worse. This is due to the fact that the estimated AoA value will be inaccurate when the size of receiving antennas becomes significant as illustrated in Fig.~\ref{fig:findAOA}.  

\subsection{Localization Results}
\begin{figure}
\centering
\includegraphics[width = 3.4 in]{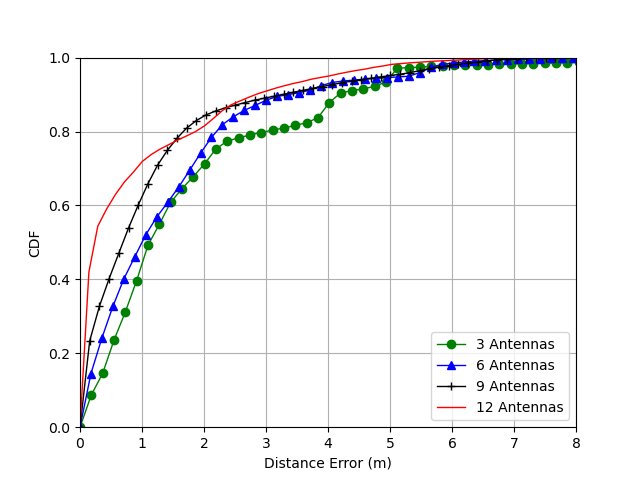}
\caption{The CDF of localization errors for different number of antennas.}
\label{fig:localization}
\end{figure}

In the second experiment, we further compare the localization accuracy in terms of RMSE for different antenna configurations. The CDF of absolute localization errors for 3, 6, 9, and 12 antennas are plotted in Fig.~\ref{fig:localization} and the overall RMSEs of localization schemes are listed in Table~\ref{tab:MSE}.

As illustrated in Fig.~\ref{fig:localization}, the localization accuracy in terms of RMSE improves when the number of receiving antennas grows, e.g. from 1.747 m for 3 antennas to 0.918 m  for 12 antennas, which is equivalent to 47.5\% improvement. From Table~\ref{tab:MSE}, we can also conclude that the absolute localization errors improves when the number of receiving antennas grows as well.

It is worth noting that, different from AoA estimation, the localization accuracy for 12 antennas is still better than that of other antenna configurations. This is because the localization estimation utilizes the amplitude information of CSI, which compensates the AoA estimation errors in the previous process. 

\begin{table} [ht]
\centering
\caption{The MAE of AoA estimation and RMSE of Localization.}
\label{tab:MSE}
\footnotesize
\renewcommand\arraystretch{1.5}
	\begin{tabular}{|c|c|c|}
	\hline  
	Numbers of Antennas&AoA errors(°)&Localization errors(m) \\ 
	\hline  
	3 antennas&5.633&1.747 \\
	\hline  
   	6 antennas&4.234&1.445 \\
	\hline  
    9 antennas&2.893&1.172 \\
	\hline  
    12 antennas&3.635&0.918 \\
	\hline	
	\end{tabular}
\end{table}

\section{Conclusion} \label{sect:conc}
In this paper, we propose an angular domain assisted localization scheme for multi-antennas configuration. With more spatial domain diversity, we propose a logistic regression based AoA estimation scheme, and substitute the AoA estimated results in the later location. Through this approach, we can achieve better localization accuracy. With the increased number of receiving antennas, e.g. from 1.746 meters with 3 antennas to 0.918 meters with 12 antennas. 

\section*{Acknowledgement}
This work was supported by the National Natural Science Foundation of China (NSFC) under Grants 62071284, 61871262, 61901251 and 61904101, the National Key Research and Development Program of China under Grants 2019YFE0196600, the Innovation Program of Shanghai Municipal Science and Technology Commission under Grant 20JC1416400, Pudong New Area Science \& Technology Development Fund, Key-Area Research and Development Program of Guangdong Province under Grants 2020B0101130012, Foshan Science and Technology Innovation Team Project under Grants FS0AA-KJ919-4402-0060 and research funds from Shanghai Institute for Advanced Communication and Data Science (SICS). 

\bibliographystyle{IEEEtran}
\bibliography{IEEEabrv,bb_rf}

\end{document}